\documentclass[letter]{ptptex}

\usepackage{graphicx}
\usepackage{wrapft}

\title{
  Separability of a Low-Momentum Effective Nucleon-Nucleon Potential
}

\author{
Hiroyuki \textsc{Kamada},$^{1,}$\footnote{E-mail: kamada@mns.kyutech.ac.jp}
 Shinichiro \textsc{Fujii},$^{2,}$\footnote{E-mail: sfujii@cns.s.u-tokyo.ac.jp
}
 Eizo \textsc{Uzu}$^{3,}$\footnote{E-mail: j-uzu@ed.noda.tus.ac.jp}
\\
 Masahiro \textsc{Yamaguchi},$^{4,5,}$\footnote{E-mail: yamagu@rcnp.osaka-u.ac.jp}
 Ryoji \textsc{Okamoto}$^{1,}$\footnote{E-mail: okamoto@mns.kyutech.ac.jp}
and Yasuro \textsc{Koike}$^{5,}$\footnote{E-mail: koike@i.hosei.ac.jp}
}

\inst{
$^1$Department  of Physics, Faculty of Engineering, Kyushu Institute of Technology,
 Kitakyushu 804-8550, Japan
\\
$^2$Center for Nuclear Study (CNS), University of Tokyo, Wako Campus of
RIKEN, Wako 351-0198, Japan
\\
$^3$Department of Physics, Faculty of Science and Technology, Tokyo University of Science,
 Chiba 278-8510, Japan
\\
$^4$Research Center for Nuclear Physics, Osaka University, Ibaraki 567-0047, Japan
\\
$^5$Science Research Center, Hosei University, 
Tokyo 102-8160, Japan
}

\recdate{October 26, 2005}

\abst{
A realistic nucleon-nucleon potential is transformed into a low-momentum effective
one (LMNN) using the Okubo theory. 
The separable potentials are converted from the LMNN with a universal separable expansion 
method and a simple Legendre expansion. Through the calculation of the triton binding energies, 
the separability for the convergence of these ranks is evaluated. It is found that there is a tendency for  
the lower momentum cutoff parameter $\Lambda$ of LMNN to gain  good separability.
 }

\begin{document}

\maketitle

\noindent{1. {\it Introduction}}~~~~~~A unified theory is regarded as an {\it integration} of some independent 
Hilbert spaces, and an effective (equivalent) theory or renormalization is regarded as 
a {\it differentiation } into the physical  space (P space) of interest 
and the remaining space (Q space). 
Theories of these two types are closely related,  and they are the main 
themes of  physics. The relation between a unified theory and an effective theory 
 can be explicitly explained by the Okubo effective theory\cite{Okubo}.
This theory is universally useful in many areas of physics. 
For example, it has a useful application  to the recently developed 
chiral perturbation theory\cite{Epelbaum} in meson-nucleon systems. The original Lagrangian 
of the nucleon ({\it N}) and pion ($\pi$) fields generates a {\it NN} interaction. 
The bare {\it NN} interaction is connected with 
the $\pi${\it NN} sector, 
we need to renormalize the sector into the effective {\it NN} interaction up to the $\pi${\it N} threshold.  

In the context of  many-nucleon systems in nuclear physics, Suzuki and Okamoto  extended
the Okubo theory to a useful scheme called the unitary model operator approach (UMOA)\cite{Suzuki,Kuo}.
The UMOA is an approach to the study of many-body systems that  considers the effective 
 interactions in a nuclear medium, which are determined by solving 
the decoupling equation between the model space and its complement space. 
When the effective theory in the sense of the Okubo theory and UMOA 
is applied to a two-body system, 
it generates a low-momentum nucleon-nucleon (LMNN)
potential by defining a model 
space (P space) and its complement (Q space).  
Bogner {\it et al.} \cite{Bogner,Kuckei} suggested that their LMNN constructed with the 
G-matrix scheme is useful in application to  many-body systems.

In order to determine the accuracy of LMNN,  one needs  
to calculate the triton and the alpha-particle binding energies,  where
the Faddeev equation or the Yakubovsky equation gives the exact solutions for the given potential.
It was concluded in  Ref. 7) 
that in the case of realistic {\it NN} forces, e.g., Nijm-I \cite{Nijm} and 
CD-Bonn\cite{Machleidt} potentials, the recommended cutoff parameter $\Lambda$, must
 at least be larger than 5 fm$^{-1}$ in order to reproduce the exact values of the 
binding energies in these systems. 
The calculation of 
the ground state energy using the LMNN for the cutoff parameter $\Lambda$ $\approx$ 2 
fm$^{-1}$ yields  a considerably more attractive result (more binding energy) than the exact value.
Variational principle is in possession a repulsion property (less binding energy) which its absolute value of 
the binding energy is less than the true one.
There could be an accidental cancellation between the attraction caused by the short cutoff parameter $\Lambda$ 
and the repulsion from the variational principle.

In addition to the above discussion, we would like to 
consider another property of the LMNN interaction. 
In general, when the Hilbert space is truncated into a small P space, the structure of the 
bases is expected to be very simple and regular. The {\it NN} interaction is expanded into a 
 separable form. This greatly reduces the necessary computational time and memory size for these 
 relatively heavy calculations in  few-body 
systems\cite{Koike,Uzu}. More precisely, we try to restrict the degrees of freedom for the continuous
 variables in the 
integral equations by introducing a separable potential. 
In the case of a three-nucleon system, the 
accuracy of the calculation has been  examined using some benchmarks. 
The separable potential has the rank of the form factors that describe the behavior of the 
potential. 
The accuracy improves as the rank becomes larger but we need small rank.
The simplicity of the P space is considered to be reflected in  the 
convergence of the rank or the separability.
In the application to few-body calculations it is interesting 
to consider whether the LMNN potential has the merit of separability.
We believe that the LMNN potential possesses good separability and that it will reduce the computational 
time  of numerical
calculations. 

In the next section, we introduce two kinds of separable expansions.
The triton binding energies are calculated using these finite rank separable potentials
in \S 3.
We would like to investigate the convergence of the rank or the separability in order to show
how the Hilbert space is effectively simplified. 
Discussion of the present result and the outlook for further developments is given in \S 4.

\noindent{2. {\it Simple separable expansion and the universal separable expansion}}~~~~~~In Ref. 7), 
the LMNN interaction was obtained using two kinds of methods. 
There, it was  numerically confirmed  that the  methods proposed by Gl\"ockle and Epelbaum \cite{Gloeckle}   
and by Suzuki and Okamoto \cite{Suzuki,Kuo} lead to the same LMNN interaction.
The LMNN potential $V(p,p')$ satisfies  the following Lippmann-Schwinger equation at energy $E$:
\begin{eqnarray}
T( p,p';E) =  V(p,p') + \int _0^{\Lambda} V(p,p'') G_0(p'',E) T(p'',p';E) {p''}^2 dp''.
\label{eq.1}
\end{eqnarray}
Here,  $T$, $V$, $G_0 $ and $\Lambda$ are the transition matrix ($t$-matrix),  the 
LMNN potential, the Green function 
of two free particles, and the cutoff parameter at low momentum, respectively.

Now, the momentum variable $p$ in the integral is replaced by another variable $x$,  defined through the 
relation
\begin{eqnarray}
p = { x+1 \over 2} \Lambda ,
\end{eqnarray}
and the potential and the $t$-matrix are expanded into the separable forms 
\begin{eqnarray}
V(p,p')~pp' \approx V^{sep}(p,p')~pp' \equiv \sum_{i,j=1} ^n g_i(x) \lambda_{i,j} g_j(x') 
\end{eqnarray}
and 
\begin{eqnarray}
T(p,p';E)~pp' \approx  \sum _{i,j} ^n g_i(x) \tau _{i,j} (E) g_j(x'),
\end{eqnarray}
where $g$ and $\lambda$ are the form factor and the coupling constant. Then, we rewrite Eq. (\ref{eq.1}) as
\begin{eqnarray}
\tau_{i,j}(E)   = \lambda_{i,j} + \sum_ {k,l} ^n  \lambda_{i,k} I_{k,l} \tau_{l,j}(E), 
\label{eq.5}
\end{eqnarray}
with 
\begin{eqnarray}
I_{k,l} \equiv { \Lambda \over 2 } \int _{-1} ^{1} g_k(x)G_0(p(x)) g_l (x) dx.
\end{eqnarray} 
Equation (\ref{eq.5}) is algebraically solved using the matrix inversion method. 
The number $n$ represents the rank of the separable expansion.

The interval of integration is finite, and the potential has no singularity.
Therefore, we conjecture that  the LMNN potential is easily expanded into simple polynomials. 
The Legendre function $P_i(x)$ can be  naturally chosen for such polynomials:
\begin{eqnarray}
g_i(x)=P_i(x)
\label{eq.7}
\end{eqnarray}
and 
\begin{eqnarray}
\lambda_{i,j} = {(2i+1)(2j+1) \over 4} \int _{-1}^1 \int_{-1} ^1 V(p,p')pp' P_i(x) P_j(x') dxdx'.
\end{eqnarray}
This expansion is not new, and 
the Hanover group has succeeded\cite{Hannover} in carrying out  
 accurate Faddeev calculations for proton-deuteron scattering by using Chebyshev polynomials.
In \S 3, we call this the simple separable expansion (SSE). 

In the present context, a well-developed separable expansion scheme has also  been introduced\cite{Koike}.
In this scheme a new form factor $g_i$ is defined as
\begin{eqnarray}
g_i (p) =\langle p | g_i \rangle \equiv \langle p | V | P _i  \rangle = \int _{-1} ^1 V(p,p'(x')) P_i (x') dx'  
\end{eqnarray}
and 
\begin{eqnarray}
V ^{USE} (p,p')~pp' \equiv \langle p| g \rangle \lambda \langle g | p' \rangle
=  \sum_{i,j}^n g_i(p) \lambda^{USE} _{i,j} g_j(p')
\end{eqnarray}
with 
\begin{eqnarray}
\left[ \lambda^{USE} _{i,j} \right]^{-1}=    \int _{-1}^1 \int_{-1} ^1 V(p,p')pp' P_i(x) P_j(x') dxdx' ,
\label{eq.11}
\end{eqnarray} 
where $[~~]^{-1}$ in Eq. (\ref{eq.11}) represents  matrix inversion.
The polynomials (Legendre functions in this case) 
are required only for the linear independence, while Eq. (\ref{eq.7}) of the SSE requires 
orthonormality. Therefore, it is understood that this expansion is  a  more general method. 
We call it the universal separable expansion (USE). 
In the case of the Faddeev calculation for  $nd$ scattering,  
 high convergence was emphasized \cite{Koike}, but
we would like to investigate the separability of the LMNN interaction by using 
the SSE and the USE.

\noindent{3. {\it Calculation of the triton  binding energies using the USE and the SSE}}~~~~~~The 
dependence of the accuracy of the LMNN potential on the cutoff momentum $\Lambda$ 
has already 
been investigated\cite{Fujii, Nogga}. We are now interested in determining how the separability 
develops when  $\Lambda$ is changed. For example, we employ the CD-Bonn  potential\cite{Machleidt}, 
 which is well known as a modern precise potential.   

We calculated the  triton binding energies using  the USE and SSE for various values of the cutoff $\Lambda$
as in Ref. 7).
For the sake of simplicity, the calculation was performed only for the 5-channel
coupled Faddeev equation. More specifically,  the potential is used only for  $^1$S$_0$ and $^3$S$_1$-$^3$D$_1$ states.  
The results are listed in Table \ref{table.1}. In the second line, the exact values obtained using the stated 
 finite values of $\Lambda$, 
calculated without the 
separable approximation, are listed. The true value (i.e., that for $\Lambda = \infty$) is --8.312 MeV. 

\begin{table}
\caption{The triton binding energies. The energies are in units of MeV. The exact values fot the stated 
finite cutoff 
$\Lambda$ appear in the second line. Here,  ``SSE" and ``USE" denote the 
simple separable expansion and the universal separable expansion.}
\label{table.1}
\begin{center}
    \begin{tabular}{cccccccccccccc}
\hline
\hline
 & & \multicolumn{2}{c}{ $\Lambda$ =~~3 fm$^{-1}$}
 & & \multicolumn{2}{c}{ $\Lambda$ =~~5 fm$^{-1}$} 
 & & \multicolumn{2}{c}{ $\Lambda$ =~~10 fm$^{-1}$ }
 & & \multicolumn{2}{c}{ $\Lambda$ =~~20 fm$^{-1}$ }   \\
 & & \multicolumn{2}{c}{ --8.532}
 & & \multicolumn{2}{c}{ --8.355} 
 & & \multicolumn{2}{c}{ --8.329}
 & & \multicolumn{2}{c}{ --8.322}   \\
 rank $n$  && SSE & USE &  & SSE & USE   &  & SSE & USE   &    & SSE &  USE  \\
\hline
$20$ && \bf--8.532 & \bf--8.532 &  & \bf--8.354 & \bf--8.355 & & --8.319 & \bf--8.329 & & --8.317 &  --8.320 \\
$18$ && \bf--8.532 & \bf--8.532 &  & --8.353 & \bf--8.355 & & --8.319 & --8.327 & & --8.308 & --8.319 \\
$16$ && \bf--8.532 & \bf--8.532 &  & --8.349 & \bf--8.354 & & --8.253 & --8.382 & & --7.762 & --7.973 \\ 
$14$ && --8.531 & \bf--8.532 &  & --8.346 & \bf--8.354 & & --7.798 & --8.305 & & --6.833 & --7.196 \\ 
$12$ && --8.526 & \bf--8.532 &  & --8.328 & --8.351 & & --6.605 & --8.203 & & --5.733 & --8.089 \\
$10$ && --8.523 & \bf--8.532 &  & --7.963 & --8.311 & & --5.645 & --7.958 & &        &        \\
$ 8$ && --8.326 & --8.509 &  & --6.730 & --8.125 & &        &        & &        &        \\
$ 6$ && --7.182 & --8.014 &  & --6.417 & --7.741 & &        &        & &        &        \\
\hline
\hline
\end{tabular}
\end{center}
\end{table}


The bold numbers in Table {\ref{table.1}  perfectly agree with the exact ones for each value of the cutoff, $\Lambda$= 3, 5
 and 10 fm$^{-1}$. 
Comparing the SSE and the USE, it is seen that the USE has  good convergence,
because in the low rank steps, the USE leads to  the corresponding exact values. 
The effective potential tends to have better separability in the lower-rank separable
form.   Lower values of $\Lambda$ result in better  separability of the LMNN interaction.  

\noindent{4. {\it  Discussion and outlook}}~~~~~~We 
calculated the triton binding energies employing  the LMNN CD-Bonn potential 
with a cutoff parameter $\Lambda$ in the unitary-transformation method of the Okubo theory.  
We find that there is the tendency that the separability improves as the value of the cutoff parameter is decreased. 
The result becomes close 
to the exact value calculated with a high-rank separable potential,
and the obtained exact value depends  on the cutoff parameter.

It is well known that the binding energy consists of  the positive expectation value from the kinetic part 
($\sim $ 50 MeV)
 and the negative expectation value from the potential ($\sim $ --60 MeV). The difference of 43 keV 
from the true value in the case of $\Lambda= $5fm$^{-1}$ represents an error of only  0.1\% error of the potential 
expectation value. 
The magnitude of a cross section in the scattering process is estimated 
the potential expectation value. Therefore, such a 0.1\% error could be negligibly small.
The Faddeev three-body scattering calculation is obtained precisely without the separable 
expansion\cite{Gloeckle2,Ishikawa,Payne}, but, in the case of four-body scattering 
great
effort is still needed to obtain precise and stable solutions. 
The separable scheme reduces the 
amount of memory and cpu time needed in the numerical computation. 
The  
contour deformation integral technique requires the analyticity  
of the form factor function to avoid logarithmic 
singularities  that  arise from the two-body $t$-matrix and the Green function in momentum
space. Most  schemes used to solve the Faddeev equation with the separable expansion method  
employ this contour-deformation technique\cite{Ebenho}.  
    
The  functional form of LMNN potential  apparently has no analyticity, and therefore  
it is not easy to apply the contour-deformation technique  to the few-body scattering problem. 
Recently, the complex energy method (CEM) was introduced \cite{Kamada}. The CEM enables us 
the calculation without an analytical  form factor, because the CEM exploits
the analytic continuation of energy. 
The solution is obtained by using the complex analytic continuation from 
sample solutions of the complex energies near the on-energy shell.
The idea of the complex energy was introduced in order to avoid 
  dangerous singularities. 

We plan to construct a  finite-rank precise separable potential from 
the LMNN interaction. The exsistence of such a potential is guaranteed 
by the good separability, as shown in the present work.
Precise calculations   using the USE potential not only for the 
three-body scattering problem but also for the four-body scattering 
problem will be carried out using the CEM.   
\\

The numerical calculations were performed mainly on a IBM  RS/6000SP 
at the Reserch Center of Nuclear Physics (Osaka University) 
in Japan, and partly 
on a Hitachi SR8000 at the Leibnitz-Rechenzentrum (die M\"unchen Hochschule) 
and a Cray SV1 at NIC (J\"ulich) in Germany.

\end{document}